\documentclass[prd,nofootinbib,aps,superscriptaddress,tightenlines,preprintnumbers]{revtex4}
\usepackage{latexsym}
\usepackage{amsmath}
\usepackage{amssymb}
\usepackage{amsfonts}
\usepackage{comment}
\usepackage{graphicx}
\usepackage{verbatim}
\usepackage{feynmp}

\newcommand{\be}{\begin{equation}}
\newcommand{\ee}{\end{equation}}
\newcommand{\bea}{\begin{eqnarray}}
\newcommand{\eea}{\end{eqnarray}}

\begin{document}

\title{Cosmic Ray Protons Illuminate Dark Matter Axions}

\author{H. Tam and Q. Yang}
\affiliation{University of Florida, Gainesville, FL 32611 }

\begin{abstract}
Cosmic ray protons propagating in a spatially-homogeneous but time-dependent field of axions or axion-like particles (ALPs) emit photons in a way that is reminiscent of Cherenkov radiation by charged particles in a preferred background.  We compute the emission rate and energy spectrum of the photons, and discuss the possibility of their detection using the Square Kilometre Array which is currently under construction.  In the case of a non-detection, constraints can be placed on the parameter space of ALPs whose mass lie between $10^{-7}$eV and $10^{-5}$ eV under the assumption that they are the primary constituent of dark matter.
\end{abstract}

\date\today
\maketitle

\section{Introduction}

We are now in an exciting era for particle physics.  In the high energy frontier, the Large Hadronic Collider has begun operation, giving us unprecedented access to physics at the TeV scale.  What has perhaps garnered less attention, but is certainly no less important, is that time may also be ripe for the discovery of new physics at the sub-eV scale.  Currently, a large number of experiments around the world are actively searching for the axion, its cousins axion-like particles (ALPs), and various light particles (e.g. chameleon, hidden photons, mini-charged particles, etc.), placing ever improving limits on their parameter spaces.  

The exploration of particle physics at the sub-eV scale was initially motivated more than thirty years ago by the proposed solution by Peccei and Quinn to the strong CP Problem, namely the absence of CP violation in the strong interaction \cite{Peccei:1977hh, Peccei:1977ur}.  This proposal, which predicts the existence of a new pseudoscalar particle (the axion), remains the most convincing solution to the strong CP Problem as of today.  Since then, two developments have injected further impetus into the field of low-energy particle physics.  First, it has been realized that the axion has exactly the right properties to be dark matter \cite{Preskill:1982cy, Abbott:1982af, Dine:1982ah, Ipser:1983mw}.  Secondly, spin-0 particles akin to the axion (known as ALPs) are found generically in string theory \cite{Svrcek:2006yi}.  It is thus no wonder that axions and ALPs have become something that is being sought avidly by experimentalists all over the world.  

In the first axion model by Weinberg and Wilczek, the symmetry breaking occurs at the weak scale, though this was quickly ruled out by experiments \cite{Weinberg:1977ma, Wilczek:1977pj}.  It was subsequently realized that the symmetry breaking scale could be much larger, though this implies that the axion couples extremely weakly to ordinary matter \cite{Kim:1979if, Shifman:1979if, Zhitnitsky:1980tq, Dine:1981rt}.  They have thus come to be known as ``invisible''.  Nonetheless, it has been demonstrated that their existence can be probed by exploiting their coupling to photons \cite{Sikivie:1983ip, Sikivie:1985yu, Maiani:1986md}.  This idea has been implemented in various experiments around the world.  The ADMX experiment is a realization of the concept of the axion haloscope, in which some halo axions in a magnetic field are induced to convert to microwave photons, which can then be picked up by an antenna \cite{Asztalos:2003px}.  The CERN Axion Solar Telescope \cite{Irastorza:2002jf} and the Tokyo Helioscope \cite{Minowa:1998sj} are axion helioscopes which convert axions from the Sun into X-rays in a magnetic field.  On the other hand, photon regeneration experiments (``shining-light-through-wall'') do not rely on external sources of axions \cite{Redondo:2010dp}.  Instead, (some) photons in a laser beam propagating towards an opaque wall are converted into axions in a magnetic field.  The axions travel unimpeded through the wall, behind which is an identical setup of magnets, where some axions are converted back to photons which can be detected. 

\begin{figure}[ht]
\begin{minipage}[b]{0.25\linewidth}
\centering
\includegraphics[scale=0.45]{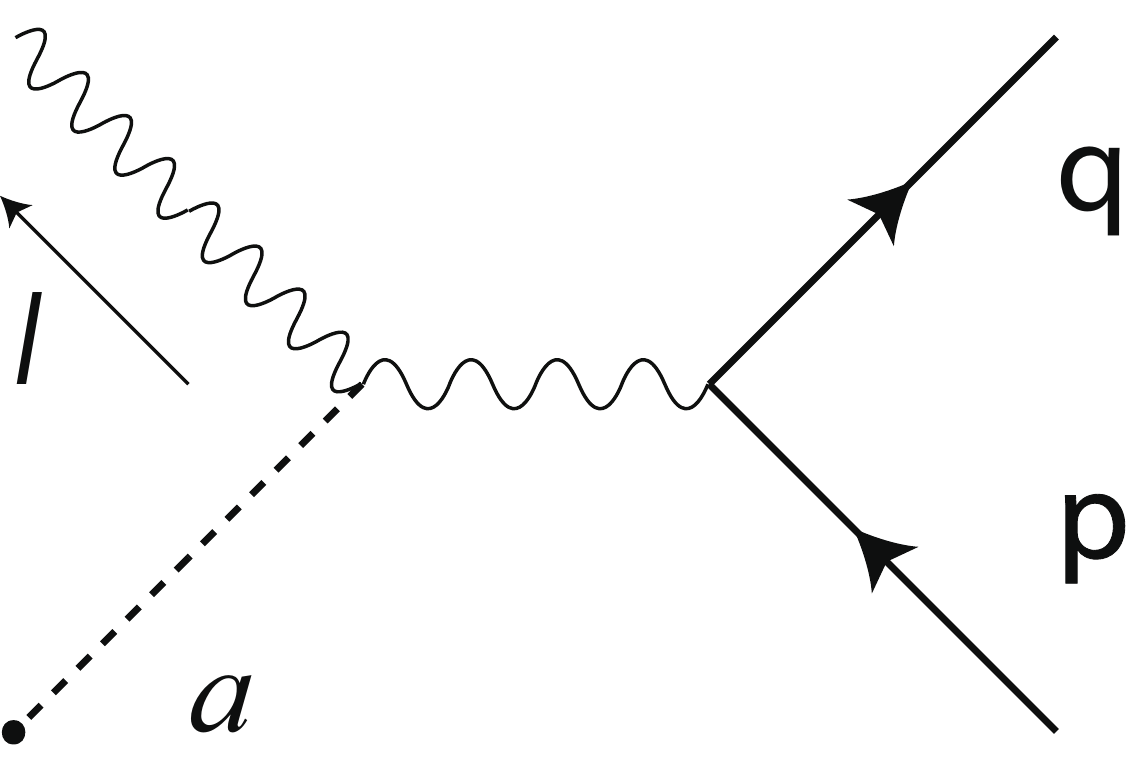}
\caption{Feynman diagram for process I}
\label{fig:feyn1}
\end{minipage}
\hspace{1.1cm}
\begin{minipage}[b]{0.25\linewidth}
\centering
\includegraphics[scale=0.45]{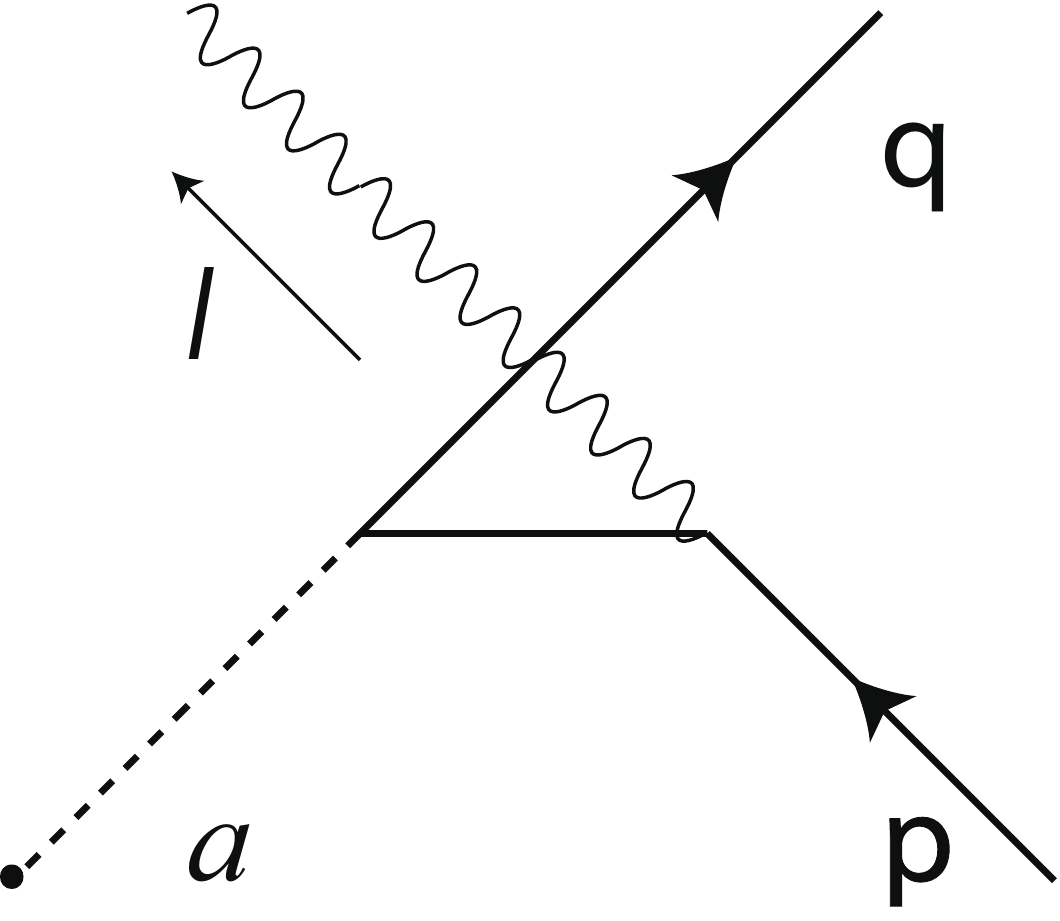}
\caption{Feynman diagram for process II}
\label{fig:feyn2}  
\end{minipage}
\hspace{0.8cm}
\begin{minipage}[b]{0.25\linewidth}
\centering
\includegraphics[scale=0.5]{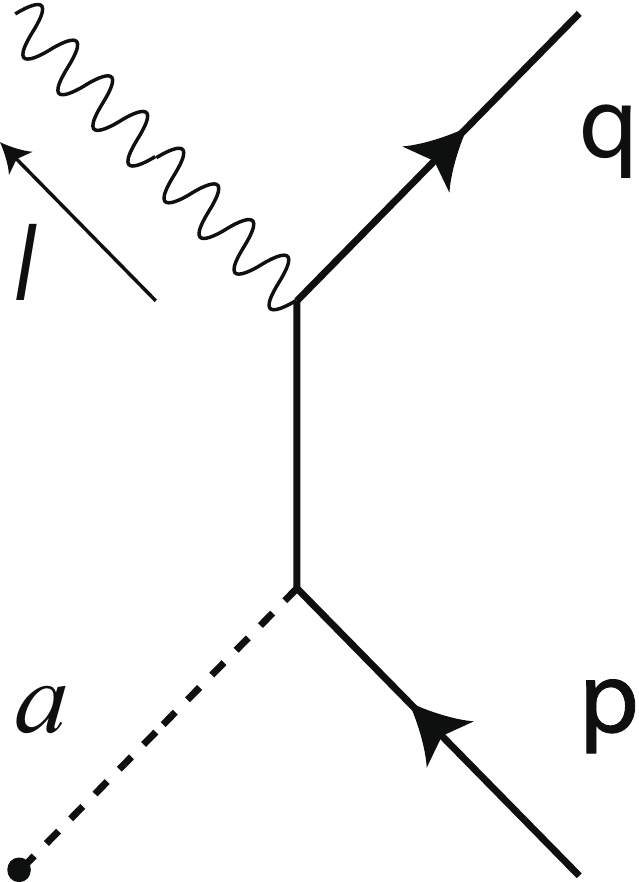}
\caption{Feynman diagram for process III}
\label{fig:feyn3} 
\end{minipage}
\end{figure}

In this paper, we propose a new observational probe of axions or ALPs, based on the fact that charged particles propagating in a time-dependent axion field emit photons.  There are three processes that contribute to the photon emission, which are shown in figures \ref{fig:feyn1}, \ref{fig:feyn2}, and \ref{fig:feyn3}.  In a sense, this phenomenon shares some resemblance with Cherenkov radiation that arises from spontaneous Lorentz violation, as the axion field effectively picks out a preferred rest frame in the universe (the one in which it is homogeneous).  The time-dependent axion field can be viewed as a source of energy (due to its time dependence), but not three-momentum (due to its homogeneity).  Consequently, some processes that are kinematically forbidden (energy and momentum conservation cannot be satisfied simultaneously) before now become possible. 

Understandably, because the axion's coupling to ordinary matter is very small, the rate of photon emission is extremely tiny.  In fact, with current technologies, performing a laboratory experiment to observe the emission is definitely not feasible.  A rough estimate shows that an electron accelerator with a reasonable length and electron flux will have to be in operation for more than the age of the universe before the emission of a single photon.  Fortunately, this phenomenon occurs naturally in cosmology: by cosmic rays (primarily protons) propagating in a time-dependent axion field.  As we will demonstrate, the abundance of axions (or ALPs) and cosmic rays in our galaxy compensate for the smallness of the coupling.  With the aid of a detector with a collecting area of $10^{10}$ cm$^2$ (for example, the Square Kilometre Array currently under construction), this can give rise to a weak but detectable signal.  

Because galactic cosmic rays also generate diffuse galactic radiation, it might appear difficult to disentangle our signal from the background, which is dominant.  Fortunately, it turns out that the energy spectrum of the photons has a well-defined peak, which is located approximately at the mass of the axion or ALP.  This is expected to be of order $\mu$eV, while diffuse galactic radiation tends to be much more energetic (GeV).  The existence of such a peak can be understood as kinematics dictates that cosmic ray protons in the low energy end of their spectrum can only produce photons whose frequency is approximately the mass of the axion or ALP.  Since both the cross section and cosmic-ray energy spectrum decrease with energy, photons whose frequency lies in the vicinity of the axion (or ALP) mass are most abundantly produced.

The propagation of charged particles in a spatially homogeneous, but time-dependent, pseudoscalar background has also been investigated in \cite{Espriu:2010bj}, in which the authors make the assumption that the time-varying background be treated as a constant in the Lagrangian (this is essentially the Lorentz-violating Chern-Simons term considered by Carroll et. al. in \cite{Carroll:1989vb}).   This additional term has the effect of modifying the dispersion relation of the photons.  As a consequence, the process of photon emission is described at leading order by a single Feynman diagram with one vertex.  According to \cite{Espriu:2010bj}, photon emission by charged particles is then possible only if the photon is massive, and that the emission angle is small.  In contrast, our calculation incorporates the time dependence of the axion field, which leads to three Feynman diagrams with two vertices each.  Using this method, we find instead that emission is possible at all angles and for massless photons. 

This paper is structured as follows.  In Section II, we review the physics behind the emission of photons by charged fermions propagating in a time-dependent pseudoscalar field.  From this, we calculate the emission rate and energy spectrum of the photons.  In Section III, we discuss the possibility of their detection with the Square Kilometre Array.  In the absence of a detection, we put constraint on the value of $f_a$.  This is followed by conclusions in Section IV.   

\section{Theoretical analysis of photon emission by protons in a pseudoscalar field}
The Lagrangian that describes the dynamics of protons ($\psi$) propagating in an axion (or ALP) field ($\phi$) is given by
\begin{equation}
\mathcal{L} = -i g_{ap}\phi \bar{\psi}\gamma^5\psi + \frac{\phi}{4}g_{a\gamma\gamma} F_{\mu\nu} \tilde{F}^{\mu\nu} + \mathcal{L}_{QED},
\end{equation}
where $\mathcal{L}_{QED}$ is the usual QED Lagrangian that describes protons and photons. The couplings $g_{app} = c_{app} m_p/f_a$ and $g_{a\gamma\gamma}=c_{a\gamma\gamma} \alpha/(2\pi f_a)$ are respectively the axion-proton and axion-photon coupling, where $c_{app}$ and $c_{a\gamma\gamma}$ are dimensional-less model-dependent parameters, typically of order unity \cite{Giannotti:2010ty}.  In this paper, we assume that $c_{app} = c_{a\gamma\gamma} = 1$. The parameter $f_a$ essentially measures the strength of the axion's coupling to the photon and proton.  For the QCD axion, $f_a$ is known as the axion decay constant, and it is constrained to $10^9 < f_a <10^{12}$ GeV by particle and nuclear physics experiments, stellar evolution, and cosmology.  This limits the QCD axion mass to \cite{Sikivie:2006ni}
\begin{equation}
m_a \approx 6 \times 10^{-6} \mbox{eV} \frac{10^{12} \mbox{GeV}}{f_a}.
\end{equation}
For ALPs, no such relation holds, and so both $f_a$ and $m_a$ are free parameters of the theory.  

The axion field is expected to oscillate, given by $\phi(t) = \phi_0 \cos(m_a t)$ \cite{Sikivie:2006ni}.  Because of its time variation and spatial homogeneity, it can be viewed as a source of energy, but not three-momentum.  This is what makes the ordinarily kinematically forbidden process of photon emission by a charged particle possible. 

\subsection{Matrix elements}
There are three processes that contribute to the emission of photons by cosmic ray photons, and their corresponding Feynman diagrams are shown in figures \ref{fig:feyn1}, \ref{fig:feyn2}, and \ref{fig:feyn3}.  The four-momenta of the incoming proton and the axion are denoted by $p^\mu$ and $a^\mu$, while those of the outgoing proton and emitted photon are $q^\mu$ and $l^\mu$ (i.e. $p(p^\mu) + a(a^\mu) \rightarrow p(q^\mu) + \gamma(l^\mu)$).  In the following, we adopt the $(+---)$ metric convention.  

The matrix element for the process (``Process I''; figure \ref{fig:feyn1}) in which a cosmic ray proton emits a virtual photon, which then interacts with the axion field to emit a real photon, is 
\begin{eqnarray} \nonumber
i\mathcal{M}_1 &=& \frac{ie g_{a\gamma\gamma}}{(p-q)^2}\bar{u}(q)\gamma^\mu u(p){{\epsilon^\alpha}_{\mu}}^{\beta\rho} (p-q)_{\alpha} l_\beta \epsilon_{\rho}^*(l).
\end{eqnarray}

The matrix element for the process (``Process II''; figure \ref{fig:feyn2}) in which a proton, subsequent to emitting a real photon, interacts with the axion field, is 
\begin{equation}
i\mathcal{M}_2 = \frac{ieg_{ak} }{2(l \cdot p)}\epsilon^*_\mu(l)\left[\bar{u}(q)\gamma^5(2p^\mu-\gamma^\nu l_\nu \gamma^\mu) u(p) \right].
\end{equation}

The matrix element for the process (``Process III''; figure \ref{fig:feyn3}) in which a proton interacts first with the axion background, then emits a real photon, is given by
\begin{equation}
i\mathcal{M}_3 = -\frac{ieg_{ak}}{2 (l \cdot q)}\epsilon^*_\mu(l)\left[\bar{u}(q)(2q^\mu + \gamma^\mu\gamma^\nu l_\nu) \gamma^5 u(p)\right].
\end{equation}

To calculate the differential cross section, we need to first compute $|\mathcal{M}_1|^2$, $|\mathcal{M}_2|^2$, $|\mathcal{M}_3|^2$, $\mathcal{M}_1 \mathcal{M}_2^*$, $\mathcal{M}_1 \mathcal{M}_3^*$, $\mathcal{M}_2 \mathcal{M}_3^*$. 

Squaring $\mathcal{M}_1$, averaging over initial proton spins, and summing over final photon polarizations, we have
\begin{eqnarray} 
\frac{1}{2} \sum_{spins} |\mathcal{M}_1|^2 &=& -\frac{4g_{a\gamma\gamma}^2 e^2 }{ (p-q)^4} \left[4m_p^2(q\cdot l)(p \cdot l) - 4(p\cdot q)(p \cdot l)(q \cdot l) + 4m_p^2 (l \cdot a)^2 - 2(l\cdot a)^2(p \cdot q)\right].
\end{eqnarray}

Meanwhile, squaring $\mathcal{M}_2$ and $\mathcal{M}_3$ yields
\begin{equation}
\frac{1}{2} \sum_{spins} |\mathcal{M}_2|^2 = \frac{4e^2 g^2_{ak}}{(l \cdot p)^2}\left[(p \cdot l)(q \cdot l)+ m_p^2(a\cdot l) -m_p^2(p\cdot q) + m_p^4\right],
\end{equation}

and
\begin{equation}
\frac{1}{2} \sum_{spins} |\mathcal{M}_3|^2 = \frac{4e^2 g^2_{ak}}{(l \cdot q)^2}\left[(p \cdot l)(q \cdot l)+ m_p^2(a\cdot l) -m_p^2(p\cdot q) + m_p^4\right].
\end{equation}

The cross terms can likewise be straightforwardly computed:
\begin{eqnarray}
\frac{1}{2} \sum_{spins} \mathcal{M}_1\mathcal{M}_2^* &=& \frac{2im_p e^2g_{a\gamma\gamma}g_{ak} (a\cdot l)^2}{(l\cdot p)(p-q)^2} \\
\frac{1}{2} \sum_{spins} \mathcal{M}_1\mathcal{M}_3^* &=& \frac{2im_p e^2g_{a\gamma\gamma}g_{ak} (a\cdot l)^2}{(l\cdot q)(p-q)^2} \\
\frac{1}{2} \sum_{spins} \mathcal{M}_2\mathcal{M}_3^* &=& -\frac{2 e^2 g_{ak}^2  }{(l\cdot p)(l\cdot q)} \left[(l\cdot p)(l\cdot q) + (p\cdot q)(l\cdot a)-(p\cdot q)^2 + m_p^2(p \cdot q)\right].
\end{eqnarray}

\subsection{Differential cross section}
For simplicity, we evaluate the differential cross section in the rest frame of the axion field.  In this case the four-momenta are given by $p^\mu = (E_p,0,0,p), a^\mu=(m_a,\vec 0), q^\mu = (E_q, q\sin\theta,0,q\cos\theta), l^\mu = (\omega, \vec\omega)$, where $E_p = \sqrt{p^2+m_p^2}$, $E_q = \sqrt{q^2+m_p^2}$, and without loss of generality we align the z-axis with the direction of propagation of the initial proton, and restrict the scattering to the x-z plane.  $\theta$ thus denotes the angle between the direction of the emitted photon and the z-axis.  The photon's frequency $\omega$ is 
\begin{equation} \label{gammafreq}
\omega = \frac{m_a^2+2m_a E_p}{2E_p + 2m_a - 2|p|\cos\theta} .
\end{equation}

In this frame, the phase space for the final state particles is given by
\begin{equation}
\int \frac{d^3\vec q}{(2\pi)^32E_q}\frac{d^3\vec \omega}{(2\pi)^32\omega} (2\pi)^4\delta^{(4)}(p+a-q-l),
\end{equation}
which yields the differential cross section
\begin{eqnarray} \nonumber
\frac{d\sigma}{d \cos(\theta)} &=& \frac{m_a+2E_p}{16\pi (2E_p+2m_a-2|p|\cos\theta)^2\sqrt{E_p^2-m_p^2}}\left[\frac{1}{2}\sum |\mathcal{M}_1 + \mathcal{M}_2 + \mathcal{M}_3|^2\right].
\end{eqnarray}

\subsection{Emission rate of the photons}
Our galaxy is teeming with cosmic rays, whose primary constituent is protons, to which we will restrict our attention in this paper (hence $E = E_p$ below).  Our calculation of the photon emission rate is thus a conservative estimate, as other charged constituents (e.g. electrons) would also contribute to the process.  As cosmic ray protons propagate in this background time-dependent axion field, they undergo photon emission via processes described in the previous section. 

To estimate the photon flux on Earth, we make the assumption that cosmic ray protons are homogeneous and isotropic within our galaxy.  This is predicated on the observation that cosmic ray protons scatter off interstellar medium and traverse random trajectories within the Galaxy for an average of $10^7$ years in the galaxy \cite{Espriu:2010bj}.  For our calculation, we adopt the following energy spectrum for cosmic ray protons \cite{book:Stanev, Adriani:2011cu}:
\begin{equation} \label{crspectrum}
\frac{d\mathcal{F}}{dE d\Omega} = \left(\frac{3.06}{\mbox{cm}^2 \mbox{ s sr GeV}}\right) \left(\frac{E}{\mbox{GeV}}\right)^{-2.70},
\end{equation}
which we assume to hold for $E > 50 $ GeV.  Since the flux is known with less certainty for low-energy protons, we impose a cutoff in our calculation and disregard all protons with an energy below $50$ GeV as a conservative measure.  We will also neglect the contribution of extragalactic cosmic rays, which should be subdominant as compared to the galactic ones. 

Consider now a photon detector on Earth, with a field of view $\delta \Omega_d$.  With our assumptions on the distribution of cosmic rays, the detector can pick up photons originating from cosmic rays filling a region from the Earth to the edge of the Galaxy, whose boundary is determined by $\delta \Omega_d$.  Let $\mathcal{V}$ denote this region, and $d\mathcal{V} = r^2 dr \delta\Omega_d$ ($r$ is the distance from $d\mathcal{V}$ to the detector on Earth) be a differential volume element in it.  At each point in $\mathcal{V}$, there are cosmic rays propagating in all possible directions.  Locally, we define a  spherical coordinate system (with the usual coordinates $\theta$ and $\phi$), where $\theta$, as we defined earlier, denotes the angle between the velocity vector of the cosmic ray and the line connecting $d\mathcal{V}$ to the detector on Earth.  The other variable $\phi$ is the usual angle confined to the plane perpendicular to the direction of propagation of the cosmic ray. 

The number of photons emitted by the cosmic rays that fill up $d\mathcal{V}$ over an interval $dt$ is given by
\begin{equation}
dN_\gamma = n_a n_p d\mathcal{V} d\sigma \delta v dt.
\end{equation}
where $n_a$ and $n_p$ are the number density of axions and protons, $\delta v$ the velocity of the cosmic rays, and $d\sigma$ the differential cross section.  From this, we can compute the the flux of photons per unit time at the detector, so we have
\begin{eqnarray}
\frac{dN_\gamma}{dt dA} = n_a \left(\frac{dn_p}{dE} dE\right) \frac{d\sigma}{dA} \delta v d\mathcal{V},
\end{eqnarray}
where $dA = r^2 d\Omega = r^2 d\cos\theta d\phi$ is the differential area at the detector.  Using the fact that 
\begin{equation}
\frac{dn_p}{dE d\Omega} = \frac{1}{\delta v} \frac{d\mathcal{F}}{dE d\Omega},
\end{equation}
the photon flux simplifies to
\begin{equation}
\frac{dN_\gamma}{dt dA} = \left( \frac{n_a}{r^2} \frac{d\mathcal{F}_p}{dE d\Omega} \frac{d\sigma}{d\cos\theta} \right) d\mathcal{V} dE d\cos\theta.
\end{equation}

Integrating over the volume that the detector can see, the proton energy, and $\cos\theta$, and using \eqref{crspectrum} we obtain the photon flux at the detector,
\begin{equation} \label{totalrate}
\frac{dN_\gamma}{dt dA} = n_a \left(\int_{\mathcal{V}} \frac{1}{r^2}d\mathcal{V}\right)\int dE\left[\frac{3.06}{\mbox{cm}^2 \mbox{ s sr GeV}}\left(\frac{E}{\mbox{GeV}}\right)^{-2.70} \left(\int_{-1}^{1} \frac{d\sigma (E,\cos\theta)}{d\cos\theta} d\cos\theta\right)\right],
\end{equation}
which can be computed numerically.

\subsection{Energy spectrum of photons}
Due to the uncertainties inherent in the energy spectrum of low-energy cosmic rays within the Galaxy, it is not possible to determine the precise spectral shape of the emitted photons.  Nonetheless, the spectrum possesses a robust feature: it has a peak located at approximately the mass of the axion or ALP (more accurately, at $\omega_c \equiv (m_a^2 + 2m_a m_p)/(2m_p + 2m_a)$, obtained by setting $E_p = m_p$ in \eqref{gammafreq}).

\begin{figure}
\includegraphics[width=0.5\columnwidth]{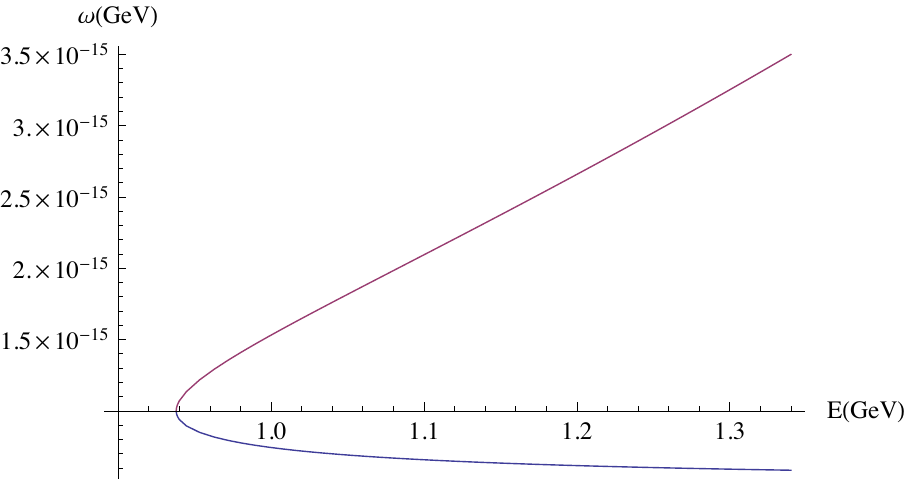}
\caption{Photon frequency $\omega$ versus the proton energy $E$, for an axion mass of $10^{-6}$ eV, according to \eqref{freqvse}.  To the right of the curve is the region in which photon generation by cosmic ray protons is allowed. The turnaround point corresponds to $\omega=\omega_c$, which is very near the axion or ALP mass.  The top (purple) and bottom (blue) curves correspond respectively to forward and backward scattering.}
\label{intbound}
\end{figure}

The existence of such a peak can be understood as follows.   From \eqref{gammafreq}, we observe that for a given $E$, the emitted photon's frequency is confined to the range $\omega_{+} \le \omega \le \omega_{-}$, where
\begin{equation} \label{freqvse}
\omega_\pm = \frac{m_a^2 + 2m_a E}{2E + 2m_a \mp 2\sqrt{E^2-m_p^2}}.
\end{equation}
The frequencies $\omega_+$ and $\omega_-$ correspond to scattering at $\theta = 0$ and $\pi$ respectively (see figure \ref{intbound}). Hence, photons of frequency $\omega$ can only be produced by cosmic ray protons with an energy larger than
\begin{equation}
E = \frac{1}{2m_a^2 - 4m_a \omega}\left(-m_a^3+3m_a^2\omega - 2m_a \omega^2 + \sqrt{-4m_p^2 m_a^2 \omega^2 + m_a^4\omega^2 + 8m_p^2 m_a\omega^3 - 4m_a^3\omega^3 + 4m_a^2\omega^4}\right).
\end{equation}
The frequency at which $dE/d\omega$ vanishes is $\omega = \omega_c$ (equivalently, $E = m_p$).  This implies that photons whose frequency is in the vicinity of $\omega_c$ are produced by cosmic rays over the widest range of energy.  More specifically, photons with a frequency near $\omega_c$ can be produced as long as the proton energy is above its rest mass.  On the other hand, photons with a very low frequency near $m_a/2$ can only be produced by enormously energetic protons which back-scatter at $\theta=\pi$, while those with a frequency $\omega \gg m_a$ can only be produced by protons whose energy exceeds approximately $\sqrt{\omega m_p^2/(2m_a)}$.  Coupled to the fact that the cosmic ray flux and cross section decrease monotonically with $E$, we thus conclude that a peak is present in the energy spectrum of the photons.  This has been verified numerically; see figure \ref{peak} for a plot of the spectrum, which features a salient peak, as expected.  

\section{Observational Consequences}
\subsection{Insufficient sensitivity to detect the QCD axion}
We numerically integrate \eqref{totalrate} and find that, for $m_a = 10^{-6}$ eV (corresponding to roughly the expected mass of the QCD axion), 
\begin{equation}
\frac{dN_\gamma}{dt dA} \sim 10^{-21} \mbox{cm}^{-2} \mbox{s}^{-1}.
\end{equation}
This estimate is quite conservative, as we only included energy $E> 50$ GeV in the integration, due to  uncertainties in the spectrum of the protons.  If we extend it to as low as, say, $10$ GeV, we gain a boost in the flux by approximately a factor of ten.  From various measurements, we know that the flux actually increases as $E$ is decreased to approximately $1.4$ GeV.  For $n_a$, we use $\rho_a/m_a$, where $\rho_a = 10^{-24}$ g/cm$^3$, the expected local halo density of the Galaxy \cite{Gates:1995dw}.  For the region $\mathcal{V}$, we adopt the value $10^{22}$ cm, which is approximately one tenth of the size of the stellar disk. The collecting area and field of view of the photon detector are taken to be that of the Square Kilometre Array: $10^{10}$ cm$^2$ and $200$ deg$^2$.  Even with such a huge surface area, we expect only one photon every $10^{11}$ s.  Clearly, it is not yet possible to detect the QCD axion through this observation.  

\begin{figure}
\includegraphics[width=0.5\columnwidth]{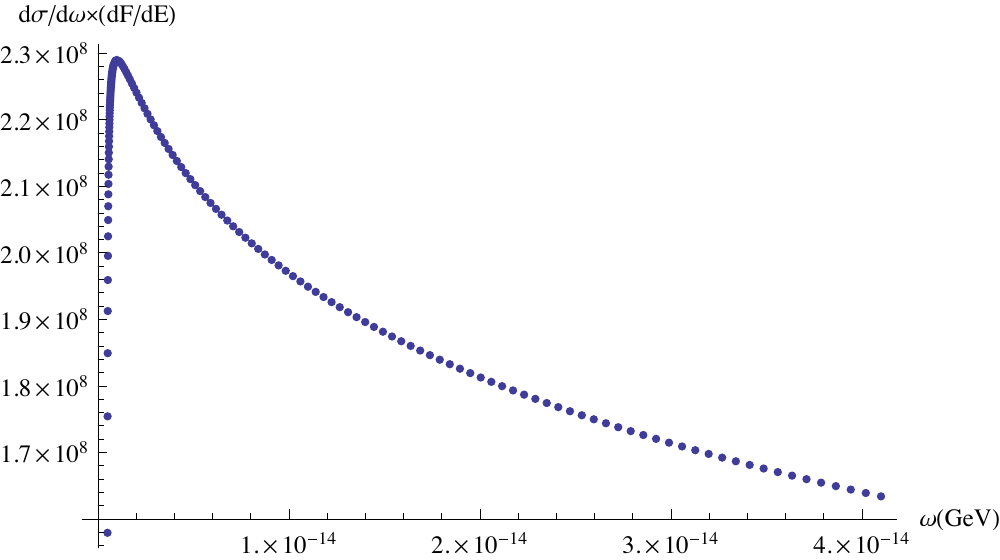}
\caption{Photon energy spectrum d$\mathcal{F}/dE \times d\sigma/d\omega$ versus photon energy $\omega$ for an axion mass of $10^{-6}$ eV, up to a normalization factor.  For $E>50$ GeV, the proton spectrum is given by $3.06 (E$/GeV)$^{-2.70}$. For $m_p<E<50$ GeV, it is chosen to be a linear function that extrapolates to zero at $E=m_p$.  This underestimates the abundance of low-energy cosmic ray protons, which however does capture the right picture that the flux vanishes at very low proton velocities.  Note that in our actual calculation we make the even-more-conservative approximation of neglecting all cosmic rays whose $E<50$ GeV. }
\label{peak}
\end{figure}

\subsection{Constraining the parameter space of ALPs}
Although the detection of the QCD axion seems out of reach given existing technologies, interesting constraint can still be placed on the ($m_a$, $f_a^{-1}$) parameter space of ALPs, under the assumption that they constitute dark matter. Their number density is thus taken to be $\rho_a/m_a$, where $\rho_a \sim 10^{-24}$ g cm$^{-3}$.  The sensitivity of the detector is assumed to be ten nanojanskys.  Over a frequency range of $10^{11}$ Hz, which is roughly what we need to see the peak in the spectrum, this translates to a minimum  rate of $10^{-3}$ photons/cm$^2$s (for the Square Kilometre Array, about $10^7$ photons per second).  Note that the constraint is only valid for an ALP whose mass lies between $2.9 \times 10^{-7}$ eV and $4.1 \times 10^{-5}$ eV, since the Square Kilometre Array can only detect photons in this range with a field of view of $200$ deg$^2$.  For higher photon energies (in the range $4.1 \times 10^{-5} $ eV $< m_a < 1.2 \times 10^{-4}$ eV that it is capable of detecting), the field of view diminishes substantially, resulting in a weaker exclusion limit.  

Our numerical result is shown in figure \ref{peak}.  The exclusion limit is slightly worse than that from laser experiments, but substantially weaker than that obtained from stellar evolution.  It is foreseeable that improvements in radio telescopy in the future could help boost the sensitivity and thus improve the constraint.  If so, our proposal would have the potential to complement existing axion experiments in the future, since it relies on very different physics from that in photon regeneration experiments (in fact, all other methods of axion search).

\begin{figure}
\includegraphics[width=0.5\columnwidth]{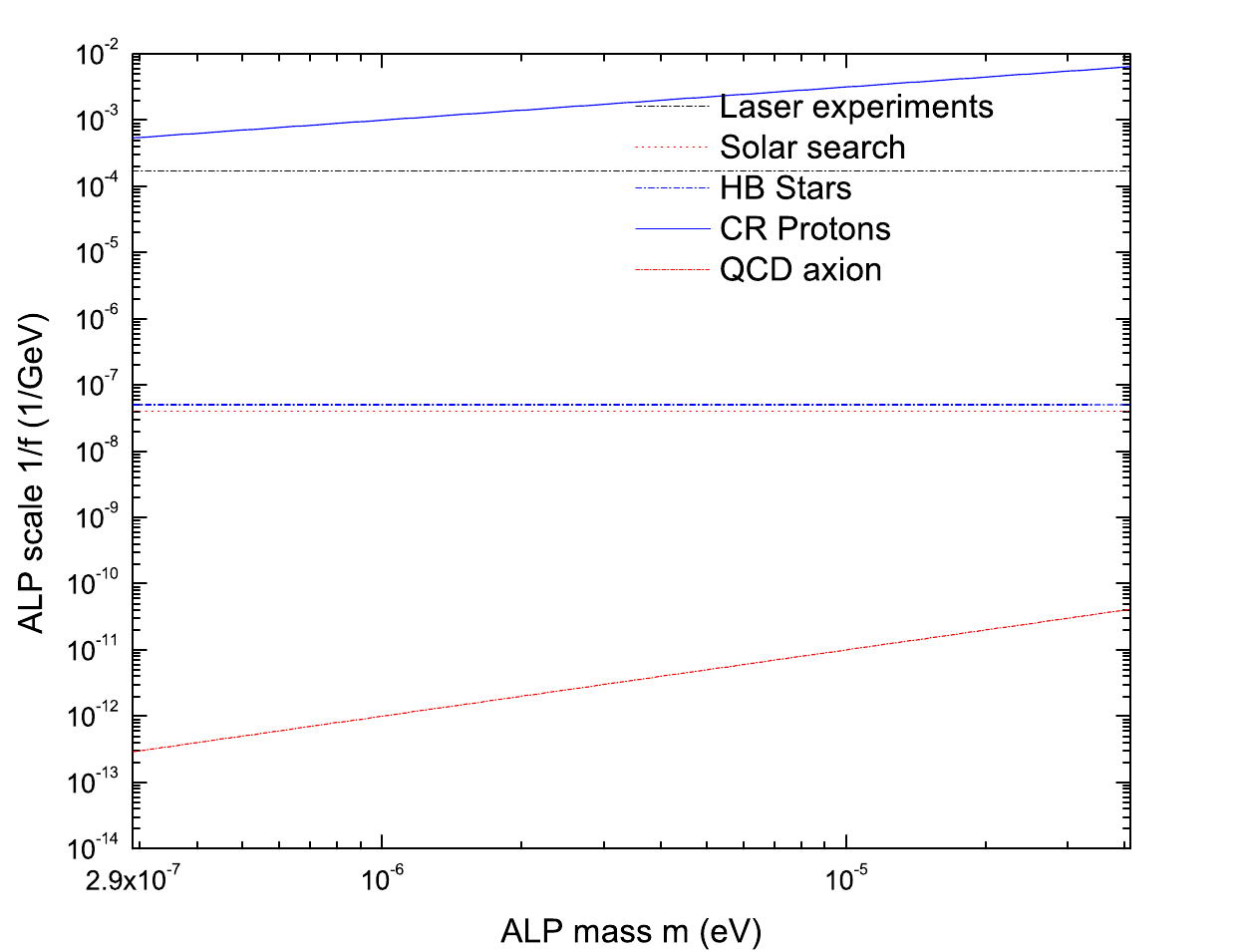}
\caption{Hypothetical exclusion limit (blue) on the mass $m_a$ and the decay constant $f_a$ for axion-like particles over mass range $2.9 \times 10^{-7}$ eV $< m_a <$ $4.1 \times 10^{-5}$ eV, in the case of a non-detection by the Square Kilometre Array.  The other limits come from laser experiments, solar searches, and stellar evolution.  The bottom line corresponds to the QCD axion.}
\label{peak}
\end{figure}

Finally, we stress that the constraint shown in figure \ref{peak} is only approximate, given the uncertainties in the halo density, and the flux spectrum and exact distribution of cosmic ray protons in the Galaxy.  Throughout our calculation, we always try to be conservative in our estimates.  For instance, we 1) disregard the contribution of all cosmic ray protons below $50$ GeV, 2) consider only proton cosmic rays, and 3) only take into account cosmic rays within a distance of $10^{22}$ cm (one-tenth of the radius of the stellar disk) from Earth.  

\section{Conclusions}
In this paper, we propose a new observational probe of axions and ALPs, based on the detection of photons emitted by cosmic ray protons which propagate in a time-dependent axion field within the Galaxy.  Since the axions couple very weakly to ordinary matter, we find, unsurprisingly, that the photon emission rate is exceedingly low, of order $10^{-21}$ cm$^{-2}$ s$^{-1}$.  Even with a detector whose collecting area is as large as $10^{10}$ cm$^2$ (e.g. the Square Kilometre Array), we expect only approximately one photon every $10^{11}$ s.  Thus detection of the QCD axion this way is out of the question at the moment. 

Nonetheless, the same mechanism can be exploited to impose exclusion limits on the parameter space of ALPs.  Since their decay constant $f_a$ can be smaller, ALPs can couple more strongly to ordinary matter, thereby increasing the rate of photon emission.  Under the assumptions that dark matter is primarily ALPs, and that a detection rate of $10^7$ photons per day is sufficient, we numerically find that a non-detection of photons by the Square Kilometre Array translates to exclusion limits on $f_a$ (for the mass range $2.9\times 10^{-7}$ eV $< m_a <  4.1\times 10^{-5}$ eV) which are slightly worse than those obtained by current photon regeneration experiments.  However, since the physics underlying our proposed observation is quite different from that used in any other axion/ALP searches, it holds the promise for serving as an excellent complement to the various axion/ALP detection experiments in the future.

\section{Acknowledgment}
We wish to thank Peter Barnes, Jim Fry, and Pierre Sikivie for insightful comments.  This work was supported in part by the U.S. Department of Energy under contract DE-FG02-97ER41029.

\end{document}